\documentclass[sigconf,final]{acmart}
\usepackage{tcolorbox}
\usepackage{enumitem}
\definecolor{grey}{gray}{0.9}
\newtcolorbox{mybox}{
    colback=grey,
    colframe=black,
    boxrule=0.2pt,
    arc=0pt,
    boxsep=0.2pt,
    left=0.2pt,
    right=0.2pt,
    top=0pt,
    bottom=0pt
}

\AtBeginDocument{%
  }

\begin{document}

\title{Root Cause Localization for Microservice Systems in Cloud-edge Collaborative Environments}


\author{Yuhan Zhu}
\affiliation{%
  \institution{Wuhan University}
  \country{China}}
\email{zhuyuhan2333@whu.edu.cn}

\author{Jian Wang}
\affiliation{%
  \institution{Wuhan University}
  \country{China}}
\email{jianwang@whu.edu.cn}

\author{Bing Li}
\affiliation{%
  \institution{Wuhan University}
  \country{China}}
\email{bingli@whu.edu.cn}

\author{Xuxian Tang}
\affiliation{%
  \institution{Wuhan University}
  \country{China}}
\email{2019302110298@whu.edu.cn}

\author{Hao Li}
\affiliation{%
  \institution{Wuhan University}
  \country{China}}
\email{2020302111283@whu.edu.cn}

\author{Neng Zhang}
\affiliation{%
  \institution{Sun Yat-sen University}
  \country{China}}
\email{zhangn279@mail.sysu.edu.cn}

\author{Yuqi Zhao}
\affiliation{%
  \institution{Central China Normal University}
  \country{China}}
\email{yuqizhao@ccnu.edu.cn}








\begin{abstract}
  With the development of cloud-native technologies, microservice-based software systems face challenges in accurately localizing root causes when failures occur. Additionally, the cloud-edge collaborative environment introduces more difficulties, such as unstable networks and high latency across network segments. Accurately identifying the root cause of microservices in a cloud-edge collaborative environment has thus become an urgent problem. In this paper, we propose MicroCERCL, a novel approach that pinpoints root causes at the kernel and application level in the cloud-edge collaborative environment. Our key insight is that failures propagate through direct invocations and indirect resource-competition dependencies in a cloud-edge collaborative environment characterized by instability and high latency. This will become more complex in the hybrid deployment that simultaneously involves multiple microservice systems. Leveraging this insight, we extract valid contents from kernel-level logs to prioritize localizing the kernel-level root cause. Moreover, we construct a heterogeneous dynamic topology stack and train a graph neural network model to accurately localize the application-level root cause without relying on historical data. Notably, we released the first benchmark hybrid deployment microservice system in a cloud-edge collaborative environment (the largest and most complex within our knowledge). Experiments conducted on the dataset collected from the benchmark show that MicroCERCL can accurately localize the root cause of microservice systems in such environments, significantly outperforming state-of-the-art approaches with an increase of at least 24.1\% in top-1 accuracy.
\end{abstract}





\maketitle
\section{Introduction}
The advancement of cloud-native technologies has led to the rise of microservices-based software architecture, which has become the primary way to achieve highly available and scalable software. Users interact with microservices through network communication, making latency a crucial factor in the quality of service. The development of edge computing enables the pre-deployment of microservices closer to the user, thereby enhancing access efficiency and reducing latency \cite{DBLP:journals/access/HaibehYJ22, DBLP:journals/comsur/PorambageOLYT18, DBLP:journals/iotj/KumarGGSL23}.
However, because edge services are limited by resources and computing power, it is necessary for cloud centers and edge servers to collaborate between microservices through cloud-edge collaboration techniques \cite{DBLP:journals/fgcs/NguyenPNC23, DBLP:journals/fgcs/WuPXJH23}. \par
\begin{figure}[t]
\centering
\small
\includegraphics[width=\linewidth]{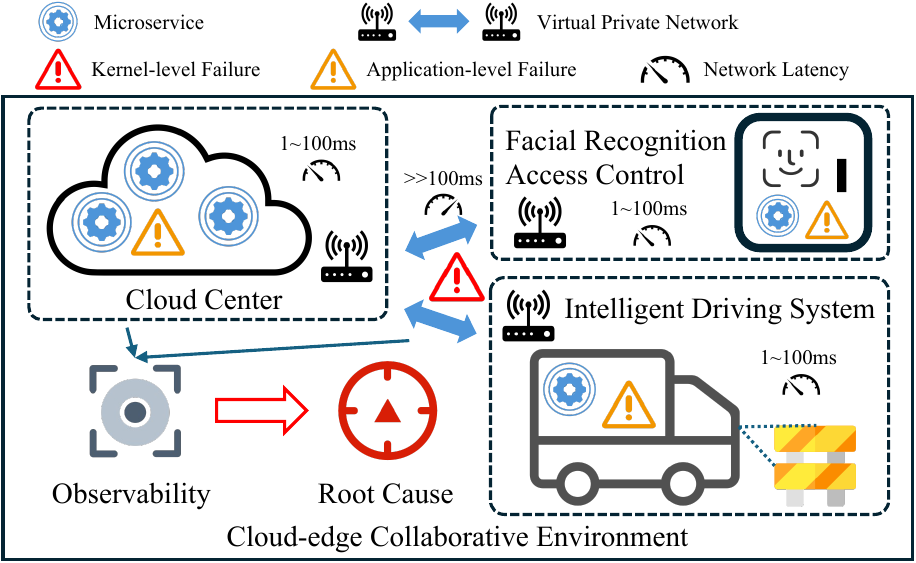}
\caption{Root cause localization in the hybrid-deployed cloud-edge collaborative environment.} \label{intro.pdf}
\end{figure}

With observability techniques \cite{sridharan_distributed_nodate}, microservices can be monitored by telemetry data, including metrics, traces, and logs, enabling Site Reliability Engineers (SREs) to better understand the performance and availability of applications. As shown in Fig. \ref{intro.pdf}, cross-segment latency is much higher than
intra-segment latency. In cloud-edge collaborative environments, instability and high latency are raised by kernel-level failures, while service dependencies add complexity to localizing application-level failures.
The root cause localization in such cloud-edge collaborative environments faces several challenges. \par
\textbf{The instability and high latency across cloud-edge network segments}: Kernel-level failures refer to failures in cloud-edge network communication, such as network disconnection or packet loss between the edge network and cloud network. Edge network usually refers to an intranet environment composed of several edge servers, and edge services provide stable services to the intranet in edge network \cite{9885783, 10190606}. However, it cannot directly access the edge network outside the intranet. The network isolation provided security and privacy for the edge network, where edge services require communication and data synchronization from cloud services when necessary. The edge network may have an unstable connection with the cloud network because of the long geographic distance, resulting in instability or high latency \cite{10.1145/3362031}.
Cloud and edge services show different features in metrics data. Prominently, cross-segment latency between the cloud and edge is much higher than intra-segment latency within the cloud or edge \cite{DBLP:conf/icws/SunZ20, DBLP:conf/icsoc/GuanLC18}. This disparity complicates the efficient discrimination between failure latency and communication noise across network segments during failures and backpropagation.
 \par
\textbf{The dynamic topology and hybrid deployment of microservice systems}: The service topology refers to the direct and indirect dependencies between microservices and foundations, which propagate application-level failure. Within a single microservice system, multiple microservices usually work together, resulting in complex and lengthy invocations that cause direct dependencies.
To cope with changes in user requests, a single microservice usually contains several instance replicas. The number of replicas can be dynamically scaled according to the required throughput \cite{DBLP:conf/www/ChakrabortyGACS23, DBLP:journals/tsc/XieWLZLH24}. Leveraging container orchestration tools, instances are deployed as containers, enabling multiple containers to run on the same physical server, thereby introducing indirect dependencies due to resource competition \cite{DBLP:journals/jnca/YuYGZH19}.
In addition to a single microservice system, the dependencies between microservice systems cannot be ignored. In real scenarios, multiple microservice systems are usually deployed in the same cluster environment, which we call hybrid deployment. In hybrid deployment, the direct and indirect dependencies between microservice systems are more complex \cite{DBLP:journals/tpds/LiZWLQWQL23, DBLP:conf/icws/ZengWLWWZ23}. 
Existing microservice root cause localization approaches \cite{DBLP:journals/csur/SoldaniB23,DBLP:conf/issre/LiuCNZZSZP19,DBLP:conf/asplos/0002LD0D21,DBLP:conf/www/YuCCGHJWSL21,DBLP:conf/sigsoft/Zhou0X0JLXH19,DBLP:conf/ccs/Du0ZS17,DBLP:conf/icse/LinZLZC16} primarily focus on scenarios involving cloud environments deploying a single microservice system with a static topology.
However, there is not yet an ideal solution for hybrid-deployed microservice systems in cloud-edge collaborative environments. \par
\textbf{The dependency on historical failure data and difficulty in implementation}: In addition to the challenges posed by deployed environments, existing deep learning-based approaches also suffer from this issue. 
Existing supervised deep learning approaches \cite{DBLP:conf/sigsoft/LiZ0LWCNCZSWDDP22, DBLP:journals/tsc/ZhangJLSZXLZMJZZP23, DBLP:conf/icse/LeeYCSL23, DBLP:conf/sigsoft/Zhang0ZSYCY22, DBLP:conf/ccs/Du0ZS17} use historical failures as training labels for models. However, the incomplete coverage of failure types in historical data results in reduced accuracy in online root cause localization. Due to the substantial amount of historical failure data, the supervised model requires extensive offline training over a long period of time. Although some unsupervised deep learning approaches \cite{DBLP:journals/jss/XinCZ23, DBLP:conf/www/ZhengCHC24, DBLP:conf/aaai/LinCWWP24} have attempted to get rid of the dependence on historical failure data, none of them can directly derive the root cause end-to-end. Instead, they train the feature representations of a causal graph or topology graph and then combine them with graph centrality computation approaches such as random walk to break down the root cause localization into two stages, increasing the difficulty in implementation. \par
Considering the above three points of issue, how to accurately localize the root cause in cloud-edge collaborative environments becomes urgent to solve. The features of failure backpropagation, along with the more complex dependencies introduced by hybrid deployment in cloud-edge collaborative environments, will be further detailed in §2.2, supported by empirical studies. \par 
In this paper, we propose MicroCERCL, an approach that accurately localizes root causes in a cloud-edge collaborative environment. Specifically, kernel-level failures concerning network communications can disrupt the entire cloud-edge collaboration, which needs to be detected first. Following this, application-level failures related to services should be localized. We model each stable topology as a heterogeneous graph. The multiple graphs created by topology changes form a heterogeneous dynamic topology stack to mine the time-series features. We combine the effects of time-series and topology features in failure backpropagation. Starting from the failure nodes detected by the anomaly detector, we train the model without relying on historical failure data, directly deriving the root cause probability of each node and forming the root cause ranking.
We conducted extensive studies on three datasets collected from the proposed benchmark. 
 MicroCERCL outperforms the baselines by 24.1\% $\sim$ 58.4\% in top-1 accuracy. Moreover, ablation studies and noise influence studies further validate the effectiveness and robustness of MicroCERCL.\par
In summary, our main contributions are summarized as follows:\begin{itemize}[leftmargin=10pt]
\item We propose MicroCERCL, a root cause localization approach in the cloud-edge collaborative environment, which can effectively and accurately localize multi-level root causes, including kernel-level and application-level failures.
\item We build a hybrid-deployed benchmark microservice system in the cloud-edge collaborative environment, which contains four widely used microservice systems. We adapt them for hybrid deployment and make them access a unified monitoring system. The entire system with complete tool chains is open-sourced on GitHub\footnote{https://github.com/WDCloudEdge/HybridCloudConfig} for further research.
\item We conduct extensive experiments on the datasets collected from the benchmark to show that MicroCERCL can accurately localize the root cause in the cloud-edge collaborative environment.  MicroCERCL shows significant improvement compared to baselines. The approach and collected datasets are also released on GitHub\footnote{https://github.com/WDCloudEdge/MicroCERCL}.
\end{itemize}
\section{Background and Motivation}
\subsection{Background}
\textbf{Observability}: Observability serves as a gauge for how accurately the internal workings of a system can be deduced from its external behavior. This concept has seen growing application in the context of microservices due to the increasing complexity of cloud-native environments and the difficulties in pinpointing the root causes. Observability relies on three types of telemetry data, including metrics, traces, and logs, to offer deeper visibility of microservice systems. Metrics reflect the operation status of microservices, microservice instances, and servers based on time-series numerical values. Traces reflect the distribution and throughput of service invocation chains. Logs contain information from the output of the kernel or business semantics of services with templated patterns.
By enhancing the ability of SREs to monitor systems effectively, observability aids in recognizing and correlating the effects within intricate causal chains, enabling the tracing back of issues to their origins. \par
\textbf{Cloud-edge collaboration}: Cloud-edge collaboration is a computing model that combines cloud computing and edge computing, aiming to take advantage of both to achieve optimal allocation of computing resources and fast response to data processing. Cloud computing is responsible for handling large-scale, complex data analysis and storage tasks, while edge computing is responsible for processing local data, reducing data transmission and latency. The combination of microservice architecture and cloud-edge collaboration can fully utilize both of their advantages. The microservice architecture can utilize edge services to achieve low-latency response while utilizing cloud services for large-scale data processing. For example, in IoT scenarios, microservices can be deployed on edge devices, where data is collected in real time and initially processed. When machine learning analysis is required, the data can be transmitted to the cloud service for processing, and the trained model can be re-downloaded to the edge device, using cloud-edge collaboration to realize collaborative work between microservices.
\subsection{Motivation}
We performed empirical studies to show the motivations of our work. The detailed experiments will be described in §4.\par
\subsubsection{Failure backpropagation in the cloud-edge collaborative environment} \par
\begin{figure}[t]
\centering
\includegraphics[width=\linewidth]{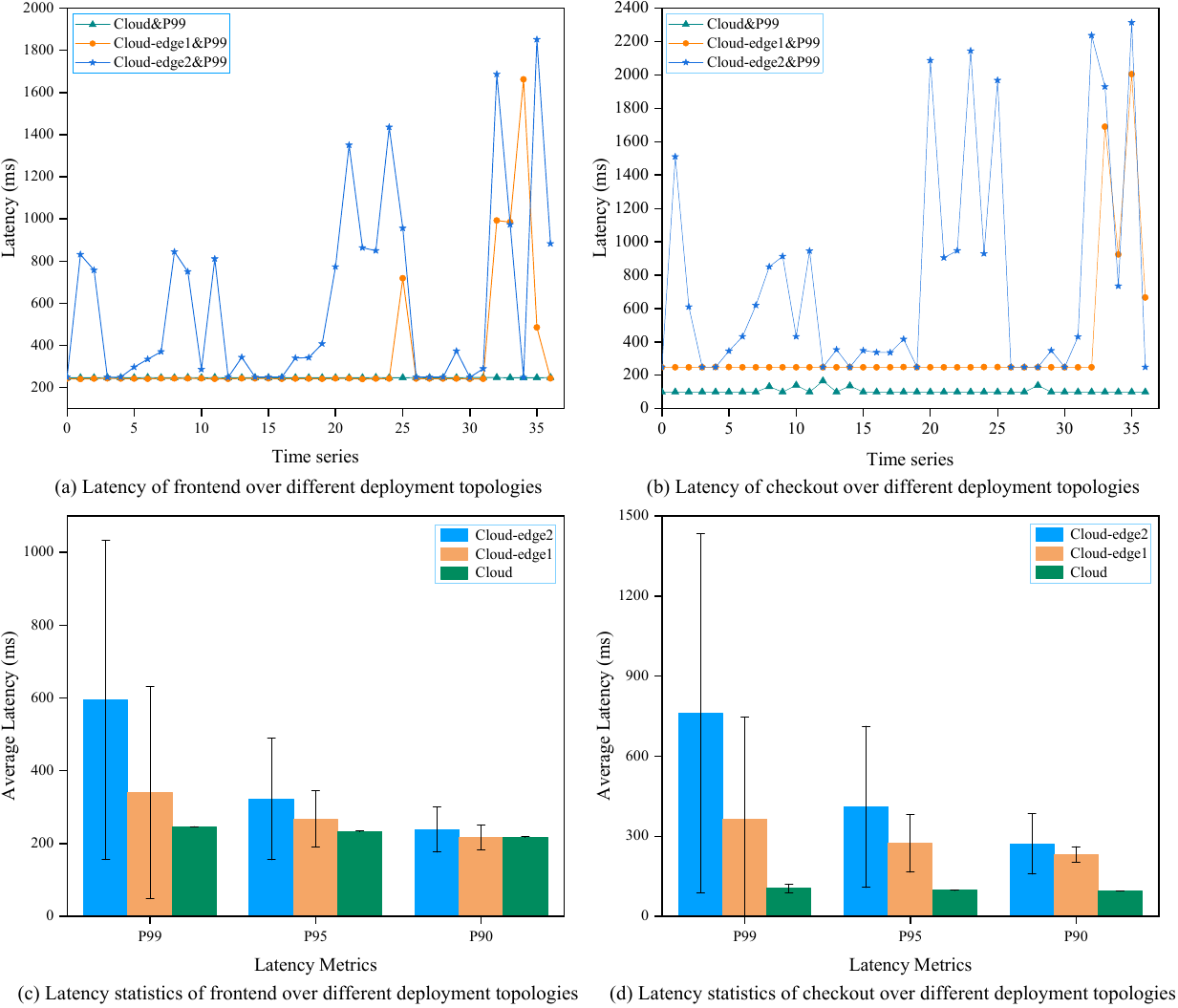}
\caption{Influence of latency over different deployment topologies.} \label{m1.pdf}
\end{figure}
\begin{figure*}[t]
\centering
\small
\includegraphics[width=\linewidth]{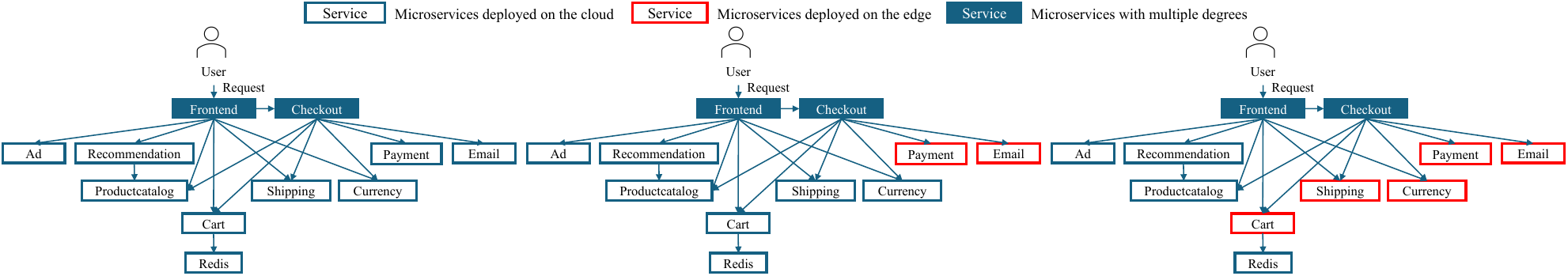}
\caption{Three topologies of service deployment in the cloud-edge collaborative environment.} \label{m1-topo.pdf}
\end{figure*}

The cloud-edge collaborative environment is different from the cloud environment because there are multiple network segments, thus introducing new types of failures in network foundations.
Besides, the cloud-edge environment serves as an important consideration for service deployment because of the high latency across network segments, thus affecting the distribution of microservice systems.
As shown in Fig. \ref{m1.pdf}, we explore the impact of different deployment topologies on service latency in the Hipster microservice system. Since extreme latency rarely occurs, we use three evaluation metrics, P99, P95, and P90, which denote the 99th, 95th, and 90th percentiles of the latency data, to better demonstrate the extreme latency that may arise from failures. The service deployment contains three topologies: cloud, cloud-edge1, and cloud-edge2, with network segments containing one cloud network segment and two edge network segments, respectively, as shown in Fig. \ref{m1-topo.pdf}. In Fig. \ref{m1.pdf}(a) and Fig. \ref{m1.pdf}(b), we find that both service latency and its fluctuation increase significantly as the number of cloud-edge co-invocations increases. This is also demonstrated in Fig. \ref{m1.pdf}(c) and Fig. \ref{m1.pdf}(d), which show the mean and standard deviation of the latency data from a statistical point of view. Also, we find that different services are affected differently by the deployment topology. The increase in service degree within the invocation topology results in a higher frequency of cross-segment communications, posing challenges to ensuring service quality. For example, the growth and fluctuation of the latency of the checkout service are stronger than those of the frontend service. The quality of service assurance needs to be coordinated with the service deployment strategy. The services should be deployed in the same network segment with a greater number of degrees, making microservice deployment in the cloud-edge collaborative environment regional. \par
\begin{mybox}
\textbf{Insight 1:} The distribution of service invocations influences the quality of service in the cloud-edge collaborative environment. The deployment of services tends to be characterized by regional concentration, making the process of failure backpropagation have the feature of topological aggregation.
\end{mybox}
\begin{figure}[t]
\centering
\small
\includegraphics[width=\linewidth]{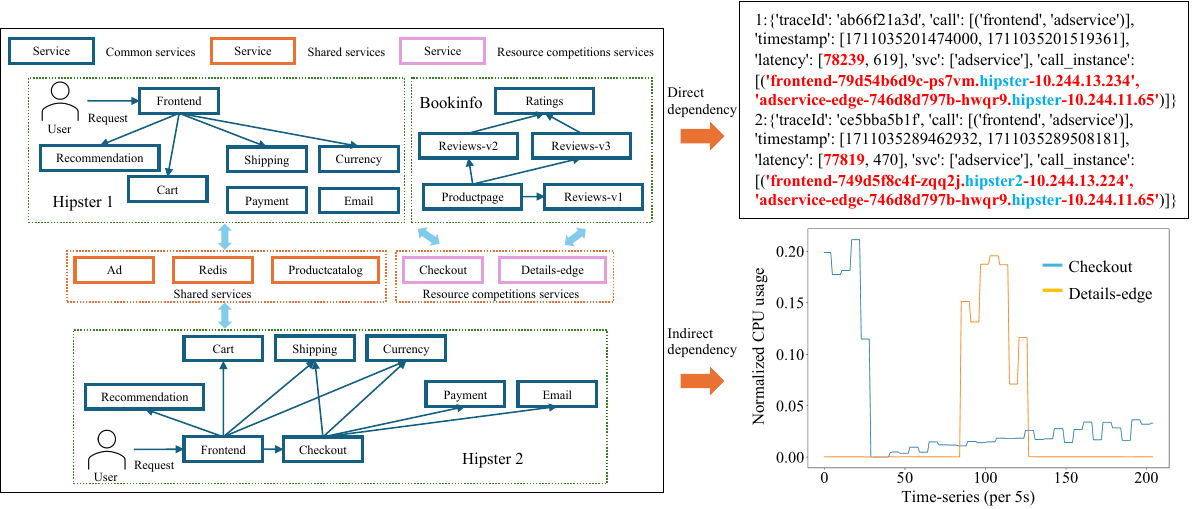}
\caption{An example of the hybrid deployment with direct and indirect dependencies.} \label{m2.pdf}
\end{figure}
\subsubsection{Root cause localization in hybrid dynamic deployment environments} Currently, open-sourced microservice systems or datasets are based on a single microservice system, which is highly functionally cohesive to provide system functionality to users. However, in real-world environments, microservice systems are usually not limited to invoking each other within the system but also between microservice systems, causing direct dependencies. Moreover, multiple microservice systems are usually hybrid-deployed in container orchestration and management tools, generating resource competition and causing indirect dependencies. Outside of the cluster, a large number of public APIs are directly accessed through the widespread use of model-as-a-service architecture, making dependencies more complex. Meanwhile, failures often occur with the creation or destruction of instances, causing topology changes.
As shown in Fig. \ref{m2.pdf}, to construct direct dependency of the above hybrid-deployed microservice systems, we perform a minimalistic environment, including two open-sourced Hipster\footnote{https://github.com/GoogleCloudPlatform/microservices-demo.git} microservice systems and modifying them to share some services. Meanwhile, deploying the services of Bookinfo\footnote{https://istio.io/latest/docs/examples/bookinfo/} and Hipster systems to the same edge server makes some of the services form an indirect dependency due to resource competition. After injecting the network delay failure, both directly associated Hipster systems observed a significant high latency in the trace data. The indirectly associated details-edge service in Bookinfo and the checkout service in Hipster also show abnormal CPU metrics after the injection of the CPU pressure failure that exacerbate resource competition. The above failures are accompanied by dynamic changes in topology.\par
\begin{mybox}
\textbf{Insight 2:} It is essential to model the hybrid deployment of microservice systems, taking into account the influence of dynamically changing topologies.
\end{mybox}
\section{Approach}
\begin{figure}[t]
\centering
\small
\includegraphics[width=\linewidth]{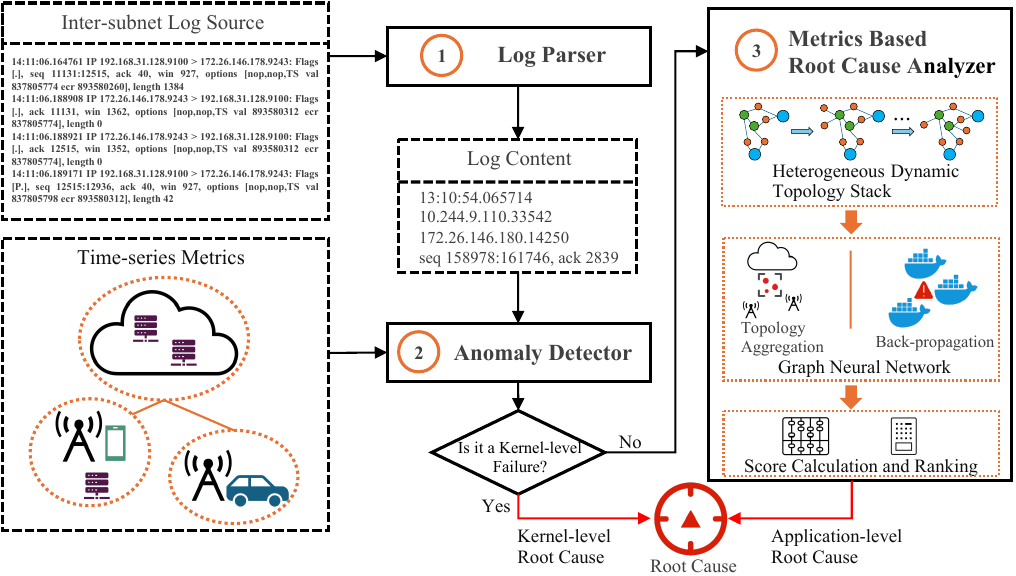}
\caption{Overall framework of MicroCERCL.} \label{Framework.pdf}
\end{figure}

As shown in Fig. \ref{Framework.pdf}, MicroCERCL consists of three main steps. In \textbf{Step 1}, we use \textbf{Log Parser} to collect cloud-edge network communication kernel logs. Then it clusters the template and extracts valid log contents. In \textbf{Step 2}, the valid contents are then input into \textbf{Anomaly Detector} to determine whether a kernel-level failure has occurred. If it is not a kernel-level failure, based on the metric data, we further detect time-series metrics to derive anomalous nodes of application-level failures. In \textbf{Step 3}, \textbf{Metrics Based Root Cause Analyzer}, we construct a heterogeneous dynamic topology stack and train a graph neural network from the anomalous nodes in the topology stack regarded as potential root causes. The loss function is jointly constructed based on the back-propagation mechanism ($bp$) and the topology aggregation mechanism ($tp$). Without any historical failure data, the probability of the root cause for each node is derived when the model converges, formatting the ranking list.

\subsection{Log Parser}
The log parser collects kernel-level logs of network communication between cloud and edge servers and further extracts contents to analyze whether a kernel-level failure has occurred. Given that the network kernel communication logs are highly templated, we extract the log templates and filter out the relevant contents. To efficiently handle the rapidly accumulating streaming logs, we use Drain \cite{DBLP:conf/icws/HeZZL17}, which maintains continuous log parsing efficiency for analysis.
Drain uses a clustering algorithm to extract the common parts of the massive logs to form template clusters, the rest of the valid content is used to validate protocol matching by the subsequent first step of the anomaly detection module to determine whether a kernel-level failure has occurred. The valid content includes network communication protocols, protocol packet formats, and packet sequences. Examples of the cluster, including source logs, templated logs, and valid contents, are shown in Fig. \ref{LogParser.pdf} (a). \par
In the process of log parsing, we found that the network kernel logs satisfy the long-tailed distribution characteristics, as shown in Fig. \ref{LogParser.pdf} (b). Kernel-level failures tend to cause common network protocols to fail to match successfully. Rather than generating new log patterns, clusters with too few samples can be pruned as noise. Based on the extracted log templates, it is also possible to filter the logs collected when subsequent failures occur, which can effectively improve the efficiency of log parsing in scenarios with high network traffic and large log volumes.\par
\subsection{Anomaly Detector}
The anomaly detector contains two steps: cloud-edge network segment anomaly detection, which can prioritize detecting kernel-level failures, and metrics based anomaly detection, which is used to detect application-level failures.\par
\begin{figure}[t]
\centering
\small
\includegraphics[width=\linewidth]{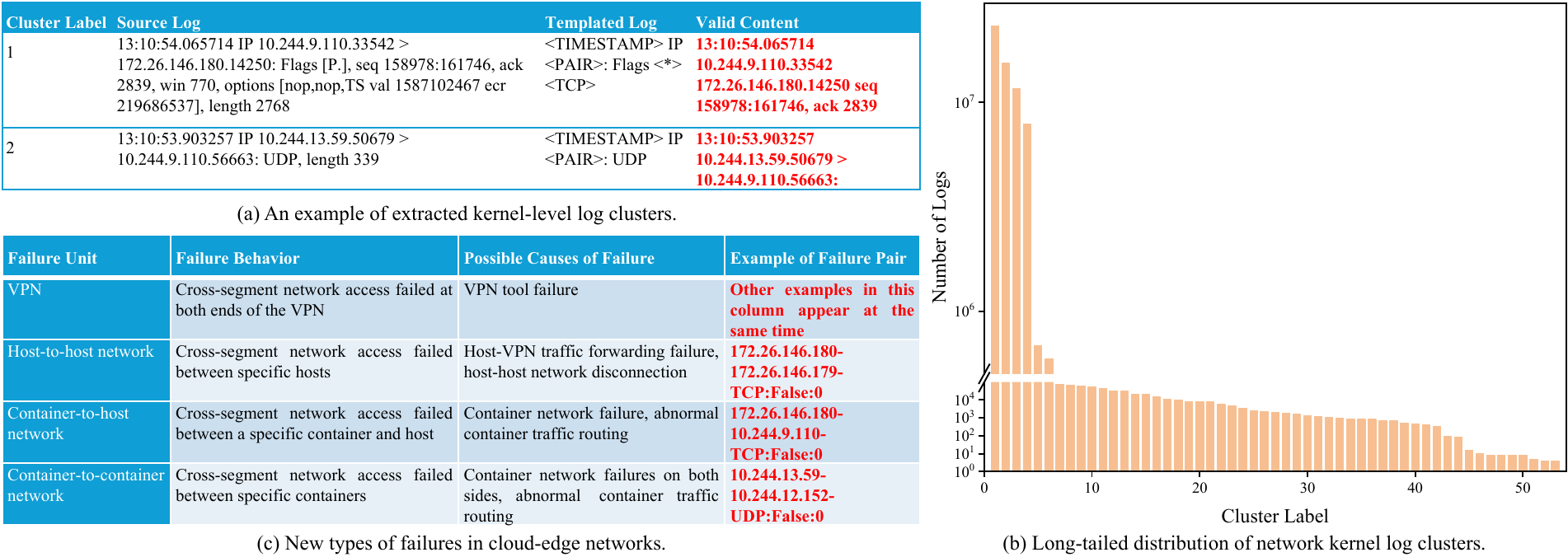}
\caption{An example of Log Parser to extract kernel-level logs with long-tailed distribution log clusters to detect new types of failures in cloud-edge networks.} \label{LogParser.pdf}
\end{figure}

\textbf{Cloud-edge network segment anomaly detection}: It detects kernel-level failures, which cause extensive paralysis between corresponding network segments. These failures need prioritization. After extracting valid log contents using a log parser, we can analyze failures based on network packet matching by protocols, considering traffic sizes from specified cross-segment host pairs, specified container-to-host pairs, and specified container pairs. These correspond to new types of failures in cloud-edge networks, as shown in Fig. \ref{LogParser.pdf} (c). We analyze the transport layer protocols, Transmission Control Protocol (TCP) and User Datagram Protocol (UDP), directly, bypassing the complexity of application layer protocols. For TCP, we match sequence packets ($seq$) and acknowledgment packets ($ack$) by packet number. Given the low bandwidth and susceptibility to network congestion of Virtual Private Networks (VPNs), we limit the Round-Trip Time (RTT) for UDP packet matching to 1 second (commonly a few to several hundred milliseconds) and use UDP request and response packets within this RTT for matching. If matching fails, it indicates a kernel-level failure, and mismatched pairs are directly identified as the root cause. Examples are shown in Fig. \ref{LogParser.pdf} (c).\par
\textbf{Metrics based anomaly detection}: Having ruled out kernel-level failures, it is necessary to further explore application-level failures. Within the selected time window $T$ of failure occurrence, $n$ metric data sampling points $\{t_1,t_2,\dots,t_n\} \in T$ are divided according to the specified sampling interval time. The collected service invocation latency is used as the metric data for service $o$, denoted as $\mathbf{\Phi}_o^\mathcal{C}$; CPU usage, memory usage, and network latency from a server are used as metric data of the server $p$, denoted as $\mathbf{\Phi}_p^\mathcal{H}$; CPU usage, memory usage, and network latency from a instance container are used as the metric data of the instance $q$, denoted as $\mathbf{\Phi}_q^\mathcal{I}$. After the time-series metric data is collected, the whole feature $\boldsymbol{\Phi}\left[t_{n}\right]$ at the moment of $t_n$ ($\mathbf{\Phi} = \{ \mathbf{\Phi}^\mathcal{C} \cup \mathbf{\Phi}^\mathcal{H} \cup \mathbf{\Phi}^\mathcal{I} \}$) is normalized by L2 paradigm as the time-series feature, and derive the feature $f^{t_{n}}$.
We use the Birch clustering approach (which is widely used in previous work \cite{DBLP:conf/noms/WuTEK20, DBLP:conf/seke/ZhangLWL21, 9527007}) to derive the anomaly metrics within $f^{t_n}$. When clustering is completed with the number of clusters greater than 1, the metric are considered to be fluctuating greatly and regarded as anomalous. 
\subsection{Metrics Based Root Cause Analyzer}
\subsubsection{Modeling of the heterogeneous dynamic topology stack.}
We further construct the topology stack, which contains multiple time-series topology graphs, when the topology is dynamically changing. We crop the time window into a dynamic topology time window set $T^d=\left\{t_1^d,t_2^d,\cdots,t_l^d\right\}$ once the topology changes, where $t_l^d$ denotes the topology interval. After deriving $T^d$, we construct a heterogeneous dynamic topology stack, denoted as $\mathcal{G}=\left\{g_{t_1^d},g_{t_2^d},\cdots,g_{t_l^d}\right\}$.
In a heterogeneous dynamic topology $g_{t_{l}^{d}}=\langle\mathcal{V}, \mathcal{E}, \mathcal{D}\rangle$ of a topology interval $t_l^d$, $\mathcal{V}$ denotes the set of nodes in the graph containing services, instances, and servers. $\mathcal{E}$ denotes the set of edges in the graph, including direct dependencies between services and instances along with indirect dependencies between instances and servers. $\mathcal{D}$ denotes different network segments of servers. The segment location of each instance on its server node is retained to distinguish different segments of the cloud or edges, indicating the topology aggregation from where instances are located. \par
\subsubsection{Construction of the graph neural network.}
\textbf{Problem formulation of the graph neural network}:
After a heterogeneous dynamic topology stack is built, the root cause localization can be transformed into the problem of finding the center node on a stack of heterogeneous topologies to be the root cause. For each topology interval, the known conditions are the topology $g_{t_l^d}$ with features of each node and the set of correspond anomalous nodes derived from the results of metric data anomaly detection, denoted as $\mathcal{A}$.
The probability of nodes within $g_{t_l^d}$ is denoted as $R_{\mathcal{V}}$.
The feature of the node $\nu_m$ derived from the time-series metric at the moment $t_n$ is denoted as $f_{\nu_m}^{t_n}$. We construct a graph neural network to calculate the probability of nodes by abstracting this problem as a multi-classification problem from the node feature $f_{\nu_m}^{t_n}$ in every topology $g_{t_l^d}$. Finally, the probabilities output from the graph neural network of the same nodes in the stack $\mathcal{G}$ are accumulated to obtain the root cause probability of the node throughout the anomalous time window, forming the root cause ranking list, denoted as $R$. \par
\begin{figure}[t]
\centering
\small
\includegraphics[width=\linewidth]{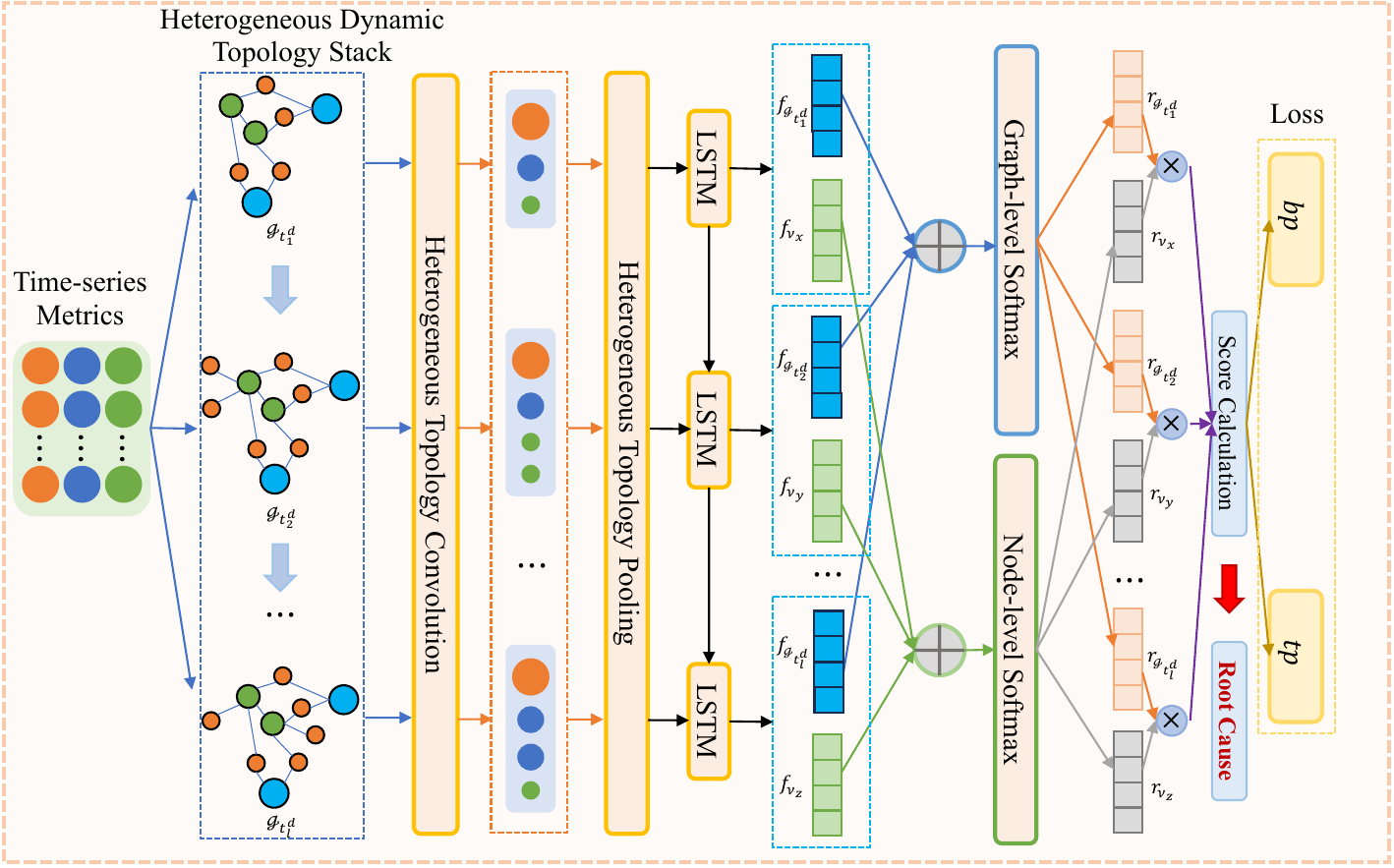}
\caption{Structure of the graph neural network.} \label{Network.pdf}
\end{figure}
\textbf{Heterogeneous topology convolution}:
For each topology $g_{t_l^d}$, to aggregate the features of heterogeneous nodes along the edges between them, we set up a different graph convolutional network for each different type of edge to extract the aggregation features. 
Each edge of node $\nu_m$ points to the neighboring node $u \in N\left(\mathcal{V}\right)$. After obtaining the neighboring features $f_u^{t_n}$ from the graph convolution network with the specified edge types, the mean vector of all the neighboring features concatenated with the original features is used as the new features of the current node. After extracting the features of $t_l^d$ at all moments, the time-series feature $f_{\nu_m}$ is mined using the LSTM network of the node $\nu_m$ in $g_{t_l^d}$. \par
\textbf{Heterogeneous topology pooling based on graph attention mechanism}:
After deriving the feature of all nodes in $g_{t_l^d}$, we stack them to be the heterogeneous topology feature $f_{g_{t_l^d}}$ of the topology interval $t_l^d$.
Since the topology structures in different topology intervals are all different, resulting in different feature dimensions. Inspired by the work \cite{DBLP:journals/corr/LiTBZ15}, we design a pooling layer based on the graph attention mechanism to pool different feature dimensions into the same dimension. We subject the heterogeneous topology feature $f_{g_{t_l^d}}$ to a node-level $Softmax$ function to derive the attention score of each node, denoted as $\Omega_\mathcal{V}$.
After that, we update the heterogeneous topology features
to expand the feature differences between nodes with the attention scores, denoted as:
\begin{equation}\small
    f_{g_{t_l^d}} = \left(\sum_{\left|t_l^d\right|}\sum_{\left|\mathcal{V}\right|}{f_{g_{t_l^d}} \times \Omega_\mathcal{V}}\right)^T \in \mathbb{R}^h
\end{equation}
$f_{g_{t_l^d}}$ denotes the pooled graph-level features with the number of neurons in the hidden layer, $h$, as the same dimension. For each node, we take the maximum of the attention scores as the root cause probability of this node, obtaining $R_\mathcal{V}=max\left(\Omega_\mathcal{V}\right)=\{{r_{\nu_1},r_{\nu_2},\cdots,r}_{\nu_m}\}$. 
\par
\textbf{Heterogeneous topology stack feature mining}:
To further fit the topology change features and time-series features within the stack, we use the LSTM network to mine the time-series features of different topology intervals,
which ultimately results in a graph-level $Softmax$ function that gives the root cause probability of each heterogeneous topology, e.g., $r_{g_{t_l^d}}$ of $g_{t_l^d}$.
The probability that node $\nu_m$ in $g_{t_l^d}$ can be denoted as $R_{\nu_m}=r_{g_{t_l^d}}\times r_{\nu_m}$. 
Finally, the probability of the same node in different heterogeneous topology graphs in this time window is obtained by accumulating. 

\textbf{Objective functions}:
The convergence objective of the graph network is to fit the anomaly detection results and cloud-edge topology aggregation that we can observe. The graph neural network model is exactly the great likelihood function that we solved to satisfy the above two features. The objective function contains two parts, including the back-propagation mechanism (\textbf{$bp$}) and the topology aggregation mechanism (\textbf{$tp$}).\par

The \textbf{$bp$} refers to the fact that after anomaly detection finds anomalous nodes $\mathcal{A}\subseteq\mathcal{V}$, these anomalous nodes are used as potential root cause nodes $R_\mathcal{A}$. For $\alpha_n\in\mathcal{A}$, since failures propagate backwards with the edges in topology, we set back-propagation parameters between each potential root cause node and its predecessor nodes, denoted as $\mathbf{W}_{bp} = \left\{ \mathbf{W}_{bp}^{R_{\alpha_1}}, \mathbf{W}_{bp}^{R_{\alpha_2}}, \cdots, \mathbf{W}_{bp}^{R_{\alpha_n}} \middle| \forall R_{\alpha_n} \in R_\mathcal{A} \right\}$. The set of probabilities of the corresponding predecessor nodes of the node $R_{\alpha_n}$ is denoted as $P _{\alpha_n}=\left\{P_{\alpha_n}^{b_1},P_{\alpha_n}^{b_2},\cdots,P_{\alpha_n}^{b_s}\right\}$. We set the target probability of $R_{\alpha_n}$ and the initial value of $\mathbf{W}_{bp}^{R_{\alpha_n}}$ to 1, assuming that each potential root cause has the greatest root cause probability, while each back-propagation has the greatest strength of 100 percent. During the training process, the $bp$ guides the root cause probability competition between multiple potential root cause nodes and their predecessor nodes. The $\mathbf{W}_{bp}$ will be adjusted and learned along with gradient descent, ultimately converging to a suitable value greater than 0, which is able to quantify the strength of the back propagation of different potential root cause nodes. This value can directly reflect the degree of causal strength between potential root cause nodes and their predecessor nodes.
The loss function of $bp$ adopts the mean squared error (MSE) between the predicted probability and the observation probability to make it more sensitive, expressed as an equation:
\begin{equation}\small
    \mathcal{L}_{bp}=\sum_{i=1}^{n}\left(\left\|{1-R_{\alpha_i}}\right\|_{2}^{2}+\sum_{j=1}^{s}\left\|\mathbf{W}_{bp}^{R_{\alpha_i}}-P _{\alpha_i}^{b_j}\right\|_{2}^{2}\right)
\end{equation}
Based on the $bp$, it is also easy to mine the hidden root causes. For example, if multiple potential root cause nodes have their predecessor nodes pointing to the same node, although the metric of this node do not fluctuate significantly, the network eventually converges all the scores pointing to this predecessor node to a larger probability value, thus top the root cause localization ranking.\par

The \textbf{$tp$} is proposed to meet cloud-edge topology aggregation. 
In the network segment $d\in\mathcal{D}$, for all nodes ($\left\{\nu_1^d,\nu_2^d,\cdots,\nu_m^d\right\}$) in the $d$, the topological aggregation property ${Aggr}_d$ is used to denote the degree of aggregation of service invocations within a network segment. The more service invocations are aggregated, i.e., the greater the topological aggregation, the stronger the causal relationship of all nodes within the same network segment, and the higher the degree of influence among them. Therefore, the loss function based on the $tp$ adopts the MSE as well, denoted as:
\begin{equation}\small
\mathcal{L}_{tp}=\sum_{d\in\mathcal{D}}{Aggr}_d=\sum_{d\in\mathcal{D}}\sum_{i=1}^{m}\left(\left\|\nu_m^d-\overline{\nu_m^d}\right\|_{2}^{2}\right)
\end{equation}
where $\overline{\nu_m^d}$ denotes the mean probability of all nodes in the same $d$. It makes all nodes within the same network segment more similar in terms of root cause probability. This mechanism enables back-propagation to be more focused within the network segment where the potential root-cause nodes are located. In the subsequent experimental evaluation, we also conducted ablation experiments on this mechanism (in §4.3.1), which demonstrated that the application of this mechanism can enhance the performance of the approach.\par
We combine $bp$ and $tp$ to effectively combine time-series and topological features, forming the final loss function:
\begin{equation}
    \mathcal{L}=\mathcal{L}_{bp}+\mathcal{L}_{tp}
\end{equation}
The constructed network is updated using this loss function iteratively.
Finally, we take the node probabilities directly from the output of the trained graph neural network model and sort them to derive the root cause localization result list $R$.

\section{EXPERIMENTAL EVALUATION}
In the evaluation, we conduct experiments to answer the following research questions (RQs): \par
\begin{itemize}[leftmargin=10pt]
\item \textbf{RQ1}: How effective is MicroCERCL regarding accuracy?\par
\item \textbf{RQ2}: How does MicroCERCL perform in terms of efficiency?\par
\item \textbf{RQ3}: How does the noise from hybrid-deployed microservice systems impact the accuracy of MicroCERCL?\par
\item \textbf{RQ4}: How do the choices of different hyperparameters affect the performance of MicroCERCL?
\end{itemize}
\subsection{Experiment Setup}

\subsubsection{Environment setup}
\par

\begin{figure*}[t]
\centering
\includegraphics[width=\linewidth]{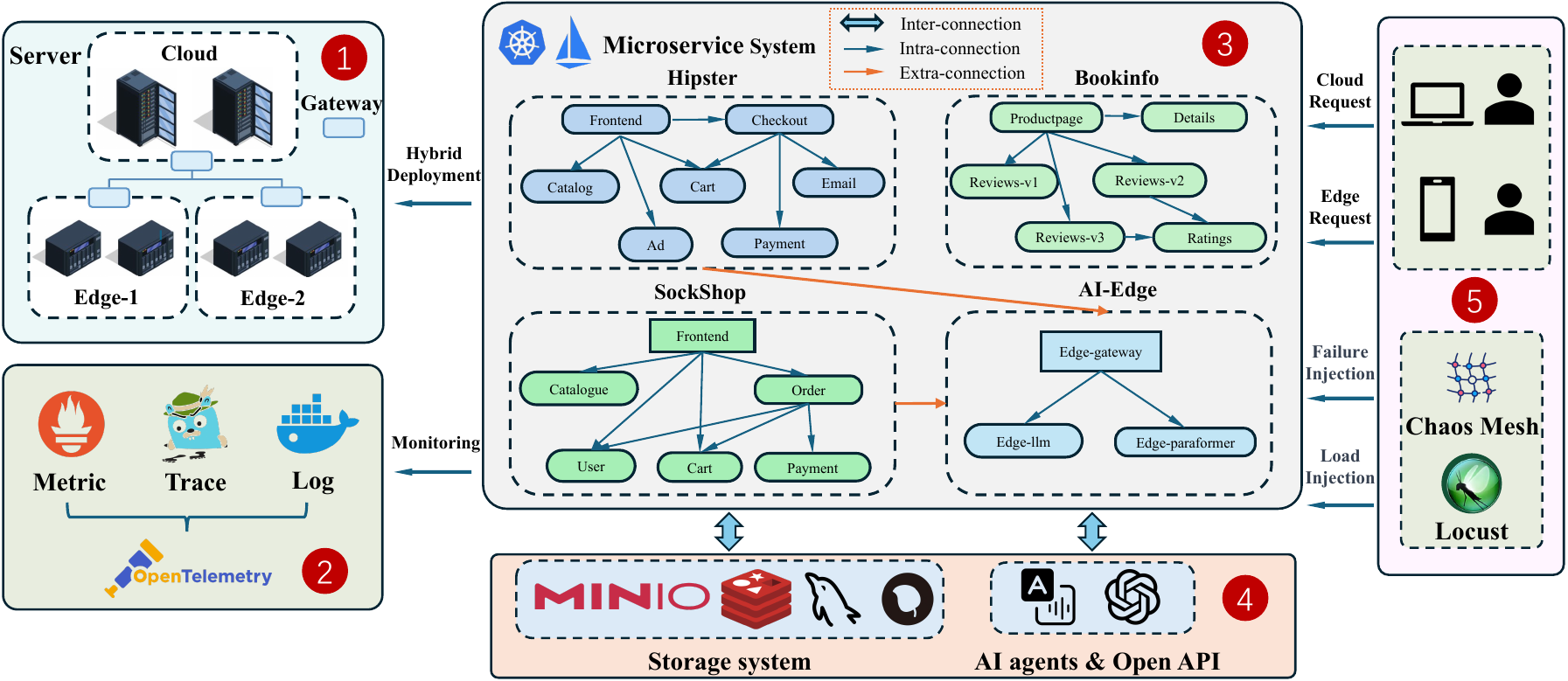}
\caption{The architecture of the hybrid-deployed benchmark microservice system in the cloud-edge collaborative environment.} \label{hybrid-cloud.pdf}
\end{figure*}

\textbf{Construction of the benchmark microservice system in a cloud-edge collaborative environment}: Currently, there is no open-source hybrid-deployed microservice system tailored for cloud-edge collaborative environments. To address this, we have developed a benchmark for large-scale hybrid-deployed microservice systems within such an environment. The architecture is shown in Fig. \ref{hybrid-cloud.pdf}.\par
The open source container orchestration and management tool Kubernetes\footnote{https://github.com/kubernetes/kubernetes.git} is selected for the cloud environment construction, and OpenYurt\footnote{https://github.com/openyurtio/openyurt.git} is an open source tool that is fully compatible with Kubernetes and realizes cloud-edge collaboration by creating a VPN channel between cloud and edge servers. The cloud-edge collaborative environment is built with four cloud servers (a master server with Intel Xeon Platinum 8369B, 8G RAM, 4vCPU, and three worker servers with Intel Xeon Platinum 8369B, 32G RAM, 4vCPU, all of them running with CentOS 7.6 OS) as cloud segments and two groups of two edge servers (each of which has an 8-core Intel i7-6700 3.40GHz CPU, 8G RAM, and runs with Ubuntu 20.04 OS) with different internal network segments as two edge segments. \par
Hybrid-deployed microservice systems are constructed with SockShop\footnote{https://github.com/microservices-demo/microservices-demo.git}, Hipster, Bookinfo, and AI-Edge. The first three have been widely used in  previous studies \cite{DBLP:conf/icse/Zhang0SZFWL022, DBLP:conf/www/YuCCGHJWSL21, DBLP:conf/icws/ChenYYZW22, DBLP:conf/sigcomm/LiCL19, DBLP:conf/issre/LiuXOJCZYMZXP20}. AI-Edge contains the AI inference service using time-consuming computational tasks, including chatbots based on the large language model and audio/video assistants based on speech-to-text algorithms. The different microservice systems within the cluster contain a total of 81 microservices. We modify them to focus on expanding the direct invocation dependencies among the microservice systems. Using Redis\footnote{https://redis.io/}, MongoDB\footnote{https://www.mongodb.com/}, Mysql\footnote{https://www.mysql.com/}, and Minio\footnote{https://github.com/minio/minio.git} as external storage, the most widely used open source systems are included for the related external storage and tools. Microservices in AI-Edge are deployed as edge services on the edge servers. All other services are randomly deployed in the cloud or edge network segments and are subject to dynamic changes.\par
\begin{table}[t]
    \centering
    \fontsize{6}{8}
    \selectfont  
    \caption{Datasets Statistics}
    \label{tab:datasetsta}
    \tabcolsep=0.12cm 
    \begin{tabular}{llll}
    \hline
        \textbf{Dataset} & \begin{tabular}[c]{@{}l@{}} \textbf{Description of} \\ \textbf{Root Cause Microservice} \end{tabular} & \begin{tabular}[c]{@{}l@{}} \textbf{Classes of} \\  \textbf{failure} \end{tabular} & \begin{tabular}[c]{@{}l@{}} \textbf{Sizes of Metric} \\ \textbf{ Sampling} \end{tabular}\\ 
    \hline
    \begin{tabular}[c]{@{}l@{}l@{}}BH \\  \\  \end{tabular} & \begin{tabular}[c]{@{}l@{}}An open-source microservice benchmark \\ showing the catalog entry of an online book store. \end{tabular} 
        & \begin{tabular}[c]{@{}l@{}l@{}}36 \\  \\  \end{tabular} & \begin{tabular}[c]{@{}l@{}l@{}}3,474,000 \\  \\ \end{tabular} \\ 
        
        \begin{tabular}[c]{@{}l@{}l@{}}HH \\  \\  \end{tabular} & \begin{tabular}[c]{@{}l@{}}An open-source web-based \\ e-commerce microservice benchmark. \end{tabular} 
        & \begin{tabular}[c]{@{}l@{}l@{}}66 \\  \\    \end{tabular} & \begin{tabular}[c]{@{}l@{}l@{}}6,369,000  \\  \\  \end{tabular}  \\
        \begin{tabular}[c]{@{}l@{}l@{}}SH \\ \\  \end{tabular} & \begin{tabular}[c]{@{}l@{}}An open-source microservice benchmark \\ for an online shop that sells socks. \end{tabular}
        & \begin{tabular}[c]{@{}l@{}l@{}}84 \\ \\  \end{tabular} & \begin{tabular}[c]{@{}l@{}l@{}}8,106,000  \\  \\ \end{tabular}  \\
        
      \hline
    \end{tabular}
\end{table}

\textbf{Load generation and failure injection}: The load generation tools named Locust\footnote{https://github.com/locustio/locust.git} are deployed with the same load ratio of different microservice systems and 100 users. Application-level failures are injected using the chaos engineering tool named ChaosMesh\footnote{https://github.com/chaos-mesh/chaos-mesh.git}, including insufficient container CPU resources, memory leakage, and network latency. For kernel-level failures, we use the Linux kernel traffic controller TC\footnote{https://man7.org/linux/man-pages/man8/tc.8.html} to inject packet loss, packet duplication, packet damage, packet disorder, network delay, and network jitter failures between the cloud-edge network communications. After load generation, a failure is injected into one of the instance replicas (which are affected by scaling up and scaling down) of service in the Bookinfo, Hipster, or SockShop within hybrid-deployed microservice systems.
\subsubsection{Dataset collection}
All hybrid-deployed microservice systems have complete access to the unified Prometheus\footnote{https://github.com/prometheus/prometheus.git} and Jaeger\footnote{https://github.com/jaegertracing/jaeger.git} monitoring tools to collect metric data and trace data, respectively. The kernel-level network logs are collected using tcpdump\footnote{https://www.tcpdump.org}, a Linux command-line packet analysis tool. After failure injection, we collect and name the corresponding dataset with the microservice system where the root cause is located, denoted as Bookinfo Hybrid (BH), Hipster Hybrid (HH), and SockShop Hybrid (SH), released on \cite{dataset}. 
We selected a time window of 5 minutes before and after the occurrence of the failure, and each failure lasts for 2-3 minutes. The sampling intervals of the time-series metric data are all 5 seconds. The parameters related to the time window facilitate the complete collection of failure fluctuation metrics. The statistics of our datasets are shown in Table \ref{tab:datasetsta}.

\subsubsection{Evaluation metrics}
To evaluate the performance of the approach experimentally, we adopted $ACC@K$ and $AVG@N$ to measure the experimental results.\par
\begin{itemize}[leftmargin=10pt]
\item
\textbf{Top-$K$ accuracy of hybrid-deployed microservice systems} ($ACC@K$) refers to the probability that the true root cause is included in the top-$K$ of the ranking list. A higher value of $ACC@K$ with a smaller $K$ indicates higher accuracy. \par
\item
\textbf{The average of top-$K$ accuracy} ($AVG@N$) is defined as the mean metric of $ACC@K$. A higher value of $AVG@N$ with a smaller $N$ indicates higher accuracy and stability.\par
\item
The \textbf{p-values} are calculated from accuracy metrics ($ACC@K$) via the $t$-test to determine statistical significance. The $t$-test offers high statistical efficacy in comparing means.
\end{itemize}
\subsubsection{Implementation and parameters setup}
We implement the prototype of MicroCERCL built on Python 3.7, PyTorch 1.13.1+cu117, and DGL 1.1.3+cu117. All experiments are conducted on a Linux server running Ubuntu 18.04 with an Intel Xeon Gold 6226R 2.90GHz CPU, 512 GB RAM, and a 24GB NVIDIA GeForce RTX 3090 GPU. We set the threshold $\beta$ to 0.07 of the Birch in the anomaly detector to accurately detect anomalous metric fluctuations and not introduce excessive noise. We set the dimension of the hidden layer in the graph neural network to 64. During model training, we used the Adaptive Moment Estimation (Adam) optimizer with an initial learning rate of 0.01 and halved every 200 epochs. We set the maximum value of the fluctuation of the loss function $\gamma$ to 1E-5, and if it stays below the maximum value for 5 consecutive epochs, the probability of each node is considered to be converged, and the model training is finished. The influence of $\beta$ and $\gamma$ are discussed in §4.3.4.
\subsection{Baseline approaches}
We use the following four metric-based state-of-the-art root cause localization approaches and two variants of MicroCERCL as baselines, all of which can be used online to localize root causes without relying on historical failure data. We modify them to adapt to the cloud-edge collaborative environment. We exclude supervised approaches due to their requirement for a substantial training dataset with labels, which is challenging to acquire in real-world scenarios.

\begin{itemize}[leftmargin=10pt]
    \item CausIL\cite{DBLP:conf/www/ChakrabortyGACS23} (2023) is a causal inference graph construction approach tailored for microservice systems, taking service instances into consideration. We implemented the approach with the random walk to localize the root cause.
    \item CausalRCA\cite{DBLP:journals/jss/XinCZ23} (2023) uses an unsupervised encoder and decoder model to train a causal graph and performs a random walk to localize the root cause in the constructed causal graph.
    \item MicroRCA\cite{DBLP:conf/noms/WuTEK20} (2020) empowers microservice topology graphs with metric data from microservices and servers and performs a personalized random walk to localize root causes.
    \item CloudRanger\cite{DBLP:conf/ccgrid/WangXMLPWC18} (2018) is a root cause localization approach based on the Pearson correlation algorithm to construct a service causal graph and perform a second-order random walk to localize root causes.
    \item MicroCERCL-A is a variant of MicroCERCL that removes the $\mathcal{L}_{tp}$ loss part without considering the topology aggregation feature in the cloud-edge collaborative environment.
    \item MicroCERCL-T is a variant of MicroCERCL that does not include LSTM to fit time-series features of metric data.
\end{itemize}\par
\begin{table*}[t]
  \centering
  \fontsize{7}{9}\selectfont  
  \caption{Effectiveness of different approaches.}
  \label{tab:accuracy}
    \begin{tabular}{lllllllllllll}
    \hline
    \begin{tabular}[c]{@{}l@{}}\textbf{Dataset}\end{tabular}&
    \begin{tabular}[c]{@{}l@{}}\textbf{Approach}\end{tabular}&
    $\textbf{ACC@1}$ & $\textbf{ACC@2}$ & $\textbf{ACC@3}$ & $\textbf{ACC@5}$ & $\textbf{ACC@10}$ & $\textbf{AVG@1}$ & $\textbf{AVG@2}$& $\textbf{AVG@3}$ & $\textbf{AVG@5}$ & $\textbf{AVG@10}$ & 
    $\textbf{p-value}$ \cr
    \hline
    \begin{tabular}[c]{@{}l@{}l@{}@{}l@{}l@{}l@{}}BH\\ \\ \\ \\ \\ \\ \\ \end{tabular} &
    \begin{tabular}[c]{@{}l@{}l@{}l@{}l@{}l@{}l@{}}CausIL \\CausalRCA\\
    CloudRanger \\MicroRCA \\MicroCERCL-A\\MicroCERCL-T\\\textbf{MicroCERCL}\end{tabular}&
    \begin{tabular}[c]{@{}l@{}l@{}@{}l@{}l@{}l@{}}0.011 \\ 0.121\\ 0.067\\ 0.145\\0.323\\0.358 \\ \textbf{0.503}\end{tabular} &
    \begin{tabular}[c]{@{}l@{}l@{}@{}l@{}l@{}l@{}}0.050 \\ 0.202\\0.270 \\0.195 \\ 0.484\\ 0.598\\\textbf{0.626}\end{tabular} &
    \begin{tabular}[c]{@{}l@{}l@{}@{}l@{}l@{}l@{}}0.112\\0.263 \\0.393 \\0.315\\0.592\\ 0.648\\ \textbf{0.687} \end{tabular} &
    \begin{tabular}[c]{@{}l@{}l@{}@{}l@{}l@{}l@{}}0.229\\0.374 \\ 0.663\\ 0.465 \\0.696\\ 0.793 \\ \textbf{0.821}\end{tabular} &
    \begin{tabular}[c]{@{}l@{}l@{}@{}l@{}l@{}l@{}}0.469 \\ 0.566\\ \textbf{0.944}\\ 0.860\\ 0.798 \\0.849\\0.911\end{tabular} &
    \begin{tabular}[c]{@{}l@{}l@{}@{}l@{}l@{}l@{}}0.011 \\ 0.121\\ 0.067\\ 0.145\\0.323 \\0.358 \\ \textbf{0.503}\end{tabular} &
    \begin{tabular}[c]{@{}l@{}l@{}@{}l@{}l@{}l@{}} 0.030 \\0.162 \\0.169 \\0.170\\ 0.403\\0.478\\ \textbf{0.565}\end{tabular} &
    \begin{tabular}[c]{@{}l@{}l@{}@{}l@{}l@{}l@{}}0.058\\ 0.195\\ 0.243\\ 0.218 \\ 0.466\\0.561\\ \textbf{0.605}\end{tabular} &
    \begin{tabular}[c]{@{}l@{}l@{}@{}l@{}l@{}l@{}} 0.114\\ 0.247\\ 0.384\\  0.304\\ 0.551\\0.650\\ \textbf{0.687}\end{tabular} &
    \begin{tabular}[c]{@{}l@{}l@{}@{}l@{}l@{}l@{}}0.239 \\ 0.371\\ 0.627\\  0.514\\ 0.655\\0.744\\ \textbf{0.784}\end{tabular} &
    \begin{tabular}[c]{@{}l@{}l@{}@{}l@{}l@{}l@{}}5.4e-6\\ 2.2e-6\\ 4.1e-3\\ 9.3e-4\\ 8.2e-5\\ 4.7e-3\\ - \end{tabular} \cr
    \hline
    \begin{tabular}[c]{@{}l@{}l@{}@{}l@{}l@{}l@{}}HH\\ \\ \\ \\ \\ \\ \\ \end{tabular} &
    \begin{tabular}[c]{@{}l@{}l@{}l@{}l@{}l@{}l@{}}CausIL\\CausalRCA\\
    CloudRanger \\MicroRCA \\MicroCERCL-A\\MicroCERCL-T\\\textbf{MicroCERCL}\end{tabular}&
    \begin{tabular}[c]{@{}l@{}l@{}@{}l@{}l@{}l@{}}0.013 \\ 0.166\\ 0.046\\ 0.488\\0.599\\0.535 \\\textbf{0.632}\end{tabular} &
    \begin{tabular}[c]{@{}l@{}l@{}@{}l@{}l@{}l@{}}0.054 \\ 0.205\\0.109 \\0.625 \\ 0.746\\0.728\\\textbf{0.756}\end{tabular} &
    \begin{tabular}[c]{@{}l@{}l@{}@{}l@{}l@{}l@{}}0.104\\0.212 \\0.195 \\0.709 \\0.789\\ 0.747\\ \textbf{0.796}\end{tabular} &
    \begin{tabular}[c]{@{}l@{}l@{}@{}l@{}l@{}l@{}}0.187\\0.232 \\ 0.362\\ 0.796\\0.829\\ 0.805 \\\textbf{0.833}\end{tabular} &
    \begin{tabular}[c]{@{}l@{}l@{}@{}l@{}l@{}l@{}}0.395 \\ 0.278\\ 0.724\\ 0.873\\ \textbf{0.896}\\0.834\\ \textbf{0.896}\end{tabular} &
    \begin{tabular}[c]{@{}l@{}l@{}@{}l@{}l@{}l@{}}0.013 \\ 0.166\\ 0.046\\ 0.488\\0.599\\0.535\\\textbf{0.632} \end{tabular} &
    \begin{tabular}[c]{@{}l@{}l@{}@{}l@{}l@{}l@{}}0.034 \\0.185 \\0.077 \\0.556 \\ 0.673\\0.617\\\textbf{0.694}\end{tabular} &
    \begin{tabular}[c]{@{}l@{}l@{}@{}l@{}l@{}l@{}} 0.057\\ 0.194\\ 0.117\\ 0.607\\ 0.711\\0.684\\\textbf{0.728} \end{tabular} &
    \begin{tabular}[c]{@{}l@{}l@{}@{}l@{}l@{}l@{}} 0.102\\ 0.209\\ 0.198\\  0.677\\ 0.755\\0.739\\\textbf{0.769}\end{tabular} &
    \begin{tabular}[c]{@{}l@{}l@{}@{}l@{}l@{}l@{}}0.210 \\ 0.236\\ 0.387\\  0.763\\ 0.816\\0.782\\\textbf{0.822}\end{tabular} &
    \begin{tabular}[c]{@{}l@{}l@{}@{}l@{}l@{}l@{}}6.1e-6\\ 3.1e-6\\ 4.5e-4\\ 2.3e-3\\ 1.5e-2\\1.3e-3\\ - \end{tabular} \cr
    \hline
    \begin{tabular}[c]{@{}l@{}l@{}@{}l@{}l@{}l@{}}SH\\ \\ \\ \\ \\ \\ \\ \end{tabular} &
    \begin{tabular}[c]{@{}l@{}l@{}l@{}l@{}l@{}l@{}}CausIL\\CausalRCA\\
    CloudRanger \\MicroRCA \\MicroCERCL-A\\MicroCERCL-T\\\textbf{MicroCERCL}\end{tabular}&
    \begin{tabular}[c]{@{}l@{}l@{}@{}l@{}l@{}l@{}}0.109 \\ 0.130\\ 0.090\\ 0.395\\0.596\\0.489\\\textbf{0.607}\end{tabular} &
    \begin{tabular}[c]{@{}l@{}l@{}@{}l@{}l@{}l@{}}0.146 \\ 0.156\\0.180 \\0.694 \\ 0.721\\0.714\\\textbf{0.732}\end{tabular} &
    \begin{tabular}[c]{@{}l@{}l@{}@{}l@{}l@{}l@{}} 0.188\\0.204 \\0.242 \\0.744 \\0.763\\0.763\\ \textbf{0.792}\end{tabular} &
    \begin{tabular}[c]{@{}l@{}l@{}@{}l@{}l@{}l@{}} 0.318\\0.242 \\ 0.370\\ 0.798 \\0.781\\0.821\\ \textbf{0.849}\end{tabular} &
    \begin{tabular}[c]{@{}l@{}l@{}@{}l@{}l@{}l@{}}0.594 \\ 0.305\\ 0.685\\ 0.882\\ 0.828\\0.855\\\textbf{0.907}\end{tabular} &
    \begin{tabular}[c]{@{}l@{}l@{}@{}l@{}l@{}l@{}}0.109 \\ 0.130\\ 0.090\\ 0.395 \\0.596\\0.489\\\textbf{0.607}\end{tabular} &
    \begin{tabular}[c]{@{}l@{}l@{}@{}l@{}l@{}l@{}}0.128 \\0.143 \\0.135 \\0.545 \\ 0.659\\0.599\\\textbf{0.672}\end{tabular} &
    \begin{tabular}[c]{@{}l@{}l@{}@{}l@{}l@{}l@{}} 0.148\\ 0.163\\ 0.171\\ 0.611 \\ 0.693\\0.678\\\textbf{0.712}\end{tabular} &
    \begin{tabular}[c]{@{}l@{}l@{}@{}l@{}l@{}l@{}} 0.200\\ 0.192\\ 0.236\\  0.680\\ 0.727\\0.732\\\textbf{0.759}\end{tabular} &
    \begin{tabular}[c]{@{}l@{}l@{}@{}l@{}l@{}l@{}}0.350 \\ 0.235\\ 0.388\\  0.763\\ 0.767\\0.793\\\textbf{0.823}\end{tabular} &
    \begin{tabular}[c]{@{}l@{}l@{}@{}l@{}l@{}l@{}}6.1e-5\\ 1.3e-6\\ 1.9e-4\\ 9.1e-3\\ 5.1e-3\\ 5.2e-3\\ - \end{tabular} \cr
    \hline
    \end{tabular}
\end{table*}
\subsection{Evaluation Results}

\subsubsection{RQ1: Effectiveness of MicroCERCL}
\textbf{Comparison with baseline approaches}:
The results of the root cause localization accuracy of the proposed MicroCERCL approach in the cloud-edge collaborative environment are shown in Table \ref{tab:accuracy}. From the results, it can be found that MicroCERCL improves significantly over other approaches by at least 24.1\%, 20.3\% and 17.3\% on $ACC@1$, $ACC@2$ and $ACC@3$ over three datasets on average. The accuracy is statistically significant from the p-value. \par
This is primarily due to the following two factors considered in MicroCERCL. First, MicroCERCL constructs a more complete heterogeneous topology stack based on service invocation data and topology structure data while considering the dynamic changes of topology and time series, which is more capable of mining the features of metric data. Second, MicroCERCL takes solving backpropagation strength as the core objective of the network, adopting an adaptive approach that is superior to approaches that use a fixed threshold to distinguish whether there is a causal relationship.
In the cloud-edge collaborative environment, due to the high cross-segment latency, the causal inference-based approaches CausIL and CausalRCA are prone to mistakenly causally correlate the high cross-segment latency with the fluctuation of failure metrics, which leads to the addition of many noise edges to the constructed causal graphs. The Pearson correlation calculation combined with the random walk-based approaches in constructing the transfer matrix is also interfered with by the instability of the cloud edge network, which leads to a decrease in the accuracy of root cause localization.

\textbf{Comparison with the variants of MicroCERCL}: 
Compared to MicroCERCL-A and MicroCERCL-T, MicroCERCL is better than them by 7.8\% and 12.3\% in terms of $ACC@1$ on average of three datasets. The accuracy is statistically significant from the p-value. To analyze the reason, it considers both the topology aggregation feature and the time-series metric feature.
Further analyzing the differences in the contribution of different datasets, we find that the topological aggregation loss function improves more significantly on the BH dataset. BH deploys more microservices of high degree in the same network segment, which leads to more significant topological aggregation. The metric data in all three datasets shows a clear temporal order, which is consistent with the typical feature of failures in metric data.\par
\subsubsection{RQ2: Efficiency of MicroCERCL}
\begin{table}[t]
  \centering
  \fontsize{7}{9}\selectfont
  \caption{Efficiency of different approaches.}
  \label{tab:effi}
  \tabcolsep=1cm
        \begin{tabular}{ll}
        \hline
        \textbf{Approach} & \textbf{Cost Time (s)} \\
        \hline
        \textbf{MicroCERCL} & \textbf{169.223} \\
        MicroRCA & 3.589 \\
        CausalRCA & 101.733 \\
        CloudRanger & 7.295 \\
        CausIL & 20.664 \\
        \hline
        \end{tabular}
\end{table}
Both the MicroCERCL approach and baselines do not rely on historical data, and the time spent is all the time from the start of triggering root cause localization to obtaining the root cause ranking. The average time overhead in all the datasets is shown in Table \ref{tab:effi}. MicroCERCL needs more time than baselines to localize each failure in 169.223 seconds. This is expected since MicroCERCL uses a more complex network with a higher iteration count to directly calculate the probability of each node. However, it is worthwhile compared to its significant accuracy improvement effect, and the efficiency of less than 3 minutes meets the timeliness requirement of root cause localization.
Compared to other non-machine learning approaches or multi-stage approaches that combine deep learning models with random walk, all of which require readjustment of multiple parameters in different environments, the applicability and generalization advantages of the MicroCERCL approach will be even more prominent.\par

\subsubsection{RQ3: Impact of the noise from the hybrid deployment.}

To explore the influence of the hybrid deployment on root cause localization, we form all possible combinations of the four hybrid-deployed microservice systems according to the number of systems ($\alpha \in \left\{4, 3, 2, 1\right\}$, e.g., when $\alpha$ = 2 in the BH dataset, the combinations contain (Bookinfo, Hipster), (Bookinfo, SockShop), and (Bookinfo, AI-Edge). The accuracy according to the $\alpha$  is shown in Table \ref{tab:dataset}.
As the number of hybrid-deployed microservice systems decreases, accuracy gradually increases due to reduced noise from metric data from non-root cause microservice systems. 
The decrease in accuracy is not statistically significant except for the HH dataset, where minimal metric fluctuations during failures increase noise confusion.
This result indicates that hybrid deployment affects root cause localization accuracy, necessitating further exploration of adaptability in such scenarios. When $\alpha = 1$, it mirrors the scenario of a single microservice system, the focus of current approaches. The accuracy maintains high overall, particularly in $ACC@1$, validating MicroCERCL's robustness.
\begin{table}[t]
  \centering
  \fontsize{7}{9}\selectfont  
  \caption{Influence of the number of hybrid-deployed microservice systems.}
  \label{tab:dataset}
    \begin{tabular}{lllllll}
    \hline
    \begin{tabular}[c]{@{}l@{}}\textbf{Dataset}\end{tabular}&
    \begin{tabular}[c]{@{}l@{}}\textbf{$\alpha$}\end{tabular}&
    $\textbf{ACC@1}$ & $\textbf{ACC@2}$ & $\textbf{ACC@3}$ & 
    $\textbf{p-value}$ \cr
    \hline
    \begin{tabular}[c]{@{}l@{}l@{}l@{}l@{}}BH\\ \\ \\ \\ \end{tabular} &
    \begin{tabular}[c]{@{}l@{}l@{}l@{}}4 \\3\\2\\ 1\end{tabular}
    &
    \begin{tabular}[c]{@{}l@{}l@{}l@{}}\textbf{0.503} \\0.579\\0.646\\ 0.676\end{tabular}
    &
    \begin{tabular}[c]{@{}l@{}l@{}l@{}}\textbf{0.626} \\0.756\\0.862\\ 0.939\end{tabular}
    &
    \begin{tabular}[c]{@{}l@{}l@{}l@{}}\textbf{0.687} \\0.816\\0.912\\ 0.989\end{tabular}
    &
    \begin{tabular}[c]{@{}l@{}l@{}@{}}-\\ 2.8e-2\\ 2.4e-2\\ 2.1e-2 \end{tabular} \cr
    \hline
    \begin{tabular}[c]{@{}l@{}l@{}l@{}l@{}}HH\\ \\ \\ \\ \end{tabular} &
    \begin{tabular}[c]{@{}l@{}l@{}l@{}}4 \\3\\2\\ 1\end{tabular}
    &
    \begin{tabular}[c]{@{}l@{}l@{}l@{}}\textbf{0.632} \\0.668\\0.750\\ 0.779\end{tabular}
    &
    \begin{tabular}[c]{@{}l@{}l@{}l@{}}\textbf{0.756} \\0.789\\0.863\\ 0.926\end{tabular}
    &
    \begin{tabular}[c]{@{}l@{}l@{}l@{}}\textbf{0.796} \\0.830\\0.905\\ 0.950\end{tabular}
    &
    \begin{tabular}[c]{@{}l@{}l@{}@{}}-\\ 1.9e-3\\ 6.6e-4\\ 9.2e-4 \end{tabular} \cr
    \hline
    \begin{tabular}[c]{@{}l@{}l@{}l@{}l@{}}SH\\ \\ \\ \\\end{tabular} &
    \begin{tabular}[c]{@{}l@{}l@{}l@{}}4 \\3\\2\\ 1\end{tabular}
    &
    \begin{tabular}[c]{@{}l@{}l@{}l@{}}\textbf{0.607} \\0.623\\0.622\\ 0.674\end{tabular}
    &
    \begin{tabular}[c]{@{}l@{}l@{}l@{}}\textbf{0.732} \\0.739\\0.749\\ 0.799\end{tabular}
    &
    \begin{tabular}[c]{@{}l@{}l@{}l@{}}\textbf{0.792} \\0.778\\0.799\\ 0.826\end{tabular}
    &
    \begin{tabular}[c]{@{}l@{}l@{}@{}}-\\7.7e-1 \\ 5.1e-2\\ 3.6e-2 \end{tabular} \cr
    \hline
    \end{tabular}
\end{table}
\subsubsection{RQ4: Impact of hyperparameters}
        
\begin{figure}[t]
\centering
\includegraphics[width=\linewidth]{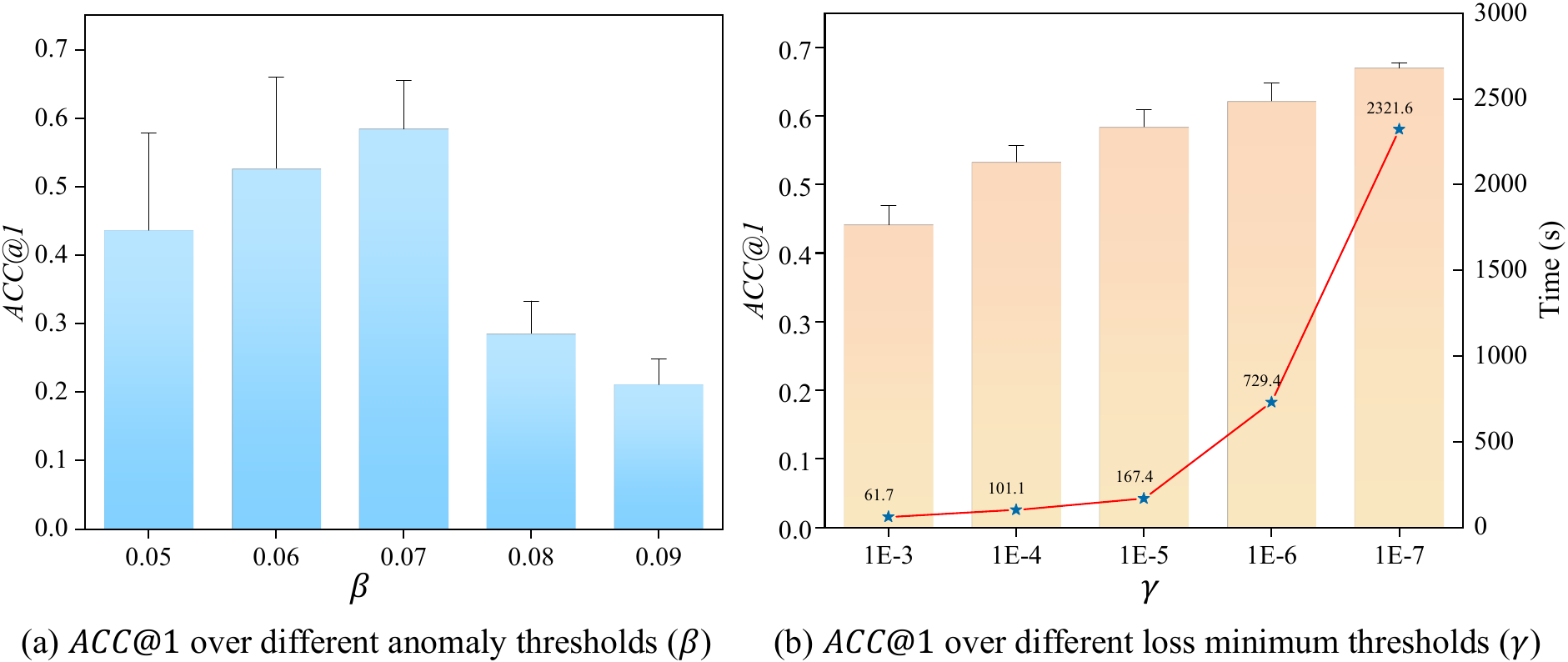}
\caption{Impact of different hyperparameters.} \label{RQ4.pdf}
\end{figure}
\textbf{The anomaly detection threshold} ($\beta$) determines the magnitude of the metric fluctuations that are considered failures. The larger the $\beta$, the more obvious degree of metric fluctuation is required, which can effectively exclude fluctuations caused by noise. However, it is also possible to ignore failures where the fluctuations are not obvious enough. Conversely, the more likely the noise will be introduced. To study the impact of the anomaly detection threshold on the root cause localization results, the $\beta$ value is set to \{0.03, 0.04, 0.05, 0.06, 0.07, 0.08, 0.09\}, and the experiment results under different $\beta$ values are shown in Fig. \ref{RQ4.pdf} (a). The accuracy reaches its highest when $\beta$ is set to 0.07, indicating that it is more capable of identifying failure metric fluctuations under this threshold while avoiding as much as possible the interference of noise.\par
\textbf{The loss minimum threshold} ($\gamma$) determines the iteration count of the model. The larger the $\gamma$, the more tolerant the probability of root-cause nodes, but the training efficiency improves. Conversely, the more accurately it takes to localize root causes, the longer the cost of time. Since the result is ultimately a sorted root cause list, it is important to ensure that the order of the list is as accurate as possible. The $\gamma$ value is set to \{1E-3, 1E-4, 1E-5, 1E-6, and 1E-7\} to analyze its impact. The effects of $\gamma$ on accuracy and corresponding efficiency are shown in Fig. \ref{RQ4.pdf} (b).
With the gradual decrease of $\gamma$, the accuracy showed a large upward trend at the beginning, and then the trend gradually decreased. The time spent on the execution also showed an upward trend, and the upward trend gradually increased. Considering the above factors, when $\gamma$ is set to 1E-5, the accuracy is higher, while the efficiency can be limited to 3 minutes.

\section{Discussion}
\subsection{Threats to Validity}
The threat to internal validity is mainly concerned with the implementation. To reduce it, we use widely used microservice systems, tools, and frameworks for cloud-edge collaboration. Meanwhile, the benchmark shield the differences between cloud or cloud-edge collaborative environments, enabling existing microservice systems to be seamlessly applied without any modifications to the source code or deployment methods. The whole chain of the benchmark and the tools involved have been carefully checked and tested.\par
The threat to external validity is mainly concerned with the generalizability. To reduce it, the experiments conducted have covered three real datasets, where microservice systems are hybrid-deployed in each dataset to reach more realistic operational scenarios, which have demonstrated the generalization. However, these datasets may not cover all microservice systems and failure types. Based on the experimental results obtained so far, especially the impact of the hybrid deployment dataset on the root cause localization accuracy obtained in §4.3.3, further demonstrated that the approach has good robustness. Meanwhile, the experiments conducted have covered a wide range of failure types faced by not only microservices but also new types of cloud-edge communication, which have never been discussed before. The approach can achieve high accuracy across a wider range of failure types.\par
\subsection{Limitation}
Although MicroCERCL achieves superior performance, there are two limitations in the current implementation. 
Firstly, in a cloud-edge collaborative environment, monitoring data reporting depends on network transmission. Network instability may result in monitoring data loss, which is not accounted for.
Secondly, when topology changes are too frequent, the excessive number of graphs in the heterogeneous dynamic topology stack can reduce the efficiency of model training, especially in large-scale topologies.

\section{Related Work}
Existing approaches can be divided into two categories: single-modal approaches \cite{DBLP:conf/www/Jiang0M023,DBLP:conf/sigsoft/HeLLZLZ18,9527007,DBLP:journals/ijseke/ZhangLWL21,DBLP:journals/tsc/ChenQH19,DBLP:conf/icws/MaLP019} and multimodal approaches based on two or more types of data \cite{DBLP:conf/kbse/HeCLYCYL22, DBLP:journals/tse/GuRR0SYLOC23}. 
\subsection{Single-modal approaches}
The single-modal approaches \cite{DBLP:conf/icse/HuoSLL23,DBLP:conf/sigsoft/LiZ0LWCNCZSWDDP22,DBLP:journals/jss/BrandonSHSPM20, DBLP:conf/sigsoft/Guo0WLJDXS20} are suitable for scenarios where multiple complete monitoring data cannot be collected while being less intrusive to the original microservice system \cite{DBLP:conf/ccgrid/YeCY21, DBLP:conf/kbse/WangWJHWKX21, DBLP:conf/icsoc/LinCZ18, DBLP:conf/sigsoft/DingZWXMWZCGGFR23, DBLP:conf/noms/WuTEK20,DBLP:conf/www/MaXWCZW20,DBLP:conf/icnsc/ZhangZZ20,DBLP:journals/pvldb/WangNLKZKB19}. DyCause \cite{DBLP:journals/tdsc/PanMJW23} extracts key information from logs of API calls, collects the metric of services from user space, and analyzes dependencies between microservices using dynamic causal inference to localize the root cause. TraceRCA \cite{DBLP:conf/iwqos/Li0JZWZWJYWCZNS21} extracts the features of anomalous trace data under the assumption that the more anomalous traces a microservice passes through, the more likely it is to be the root cause. It calculates the sum of the conditional probabilities of the microservices involved in the anomalous traces to derive the final root cause. 
CloudRanger \cite{DBLP:conf/ccgrid/WangXMLPWC18} uses the PC algorithm to construct a dynamic service causal graph from the metric data, running a second-order random walk algorithm to localize the root cause. 
DejaVu \cite{DBLP:conf/sigsoft/LiZ0LWCNCZSWDDP22} proposes an effective supervised training method by collecting multidimensional metric data and constructing a failure dependency graph to train a graph attention network \cite{velickovic_graph_2018}. DejaVu can well-fit the features of the recurring failure types, but with poor support for new failure types. CausalRCA \cite{DBLP:journals/jss/XinCZ23} utilizes the encoder and decoder combined with the latency metric to learn the causal graph structure between service nodes and then use the random walk algorithm through the causal graph to derive the root cause. However, this approach needs to judge causality based on a fixed threshold, making it difficult to construct suitable causal relationships. The drawback of these approaches is that none of them directly use kernel-level log data, which means they are unable to pinpoint kernel-level failures. \par
\subsection{Multimodal approaches}
In recent years, the fusion of multimodal monitoring data has attracted much attention in the fields of root cause localization \cite{DBLP:conf/sigsoft/YuCLCLZ23, DBLP:conf/icsoc/WhiteDFO21, DBLP:conf/kdd/ZhaoMZZ0XYFSZPL23}. Researchers have collected complete data from three modalities simultaneously through a complete monitoring system, and the data from different modalities can be complementary and correlated with each other. MicroHECL \cite{DBLP:conf/icse/LiuH0LZGLOW21} uses the correlation of failure among microservices to continuously prune irrelevant microservices to achieve performance optimization with the trace and metric data. DiagFusion \cite{DBLP:journals/tsc/ZhangJLSZXLZMJZZP23} fuses multimodal data in the data processing stage and unifies them into events. The unified information from heterogeneous modalities is provided to downstream diagnostic tasks to reduce the complexity of data processing. Unlike DiagFusion, Eadro \cite{DBLP:conf/icse/LeeYCSL23} merges the representations of each modality during the model training process and trains them together. This intermediate fusion approach allows for the integration of high-dimensional knowledge from all modalities, thus improving overall performance. MULAN \cite{DBLP:conf/www/ZhengCHC24} learns the causal structure in the microservice topology through multimodal data and then combines the random walk to locate the root cause, avoiding the dependency of historical failure data. However, current multimodal approaches all require complete monitoring tools and rely on network transmission to collect multimodal data, making them unsuitable for resource-constrained and network-unstable cloud-edge collaborative environments.\par

\section{Conclusion}
In this paper, we propose MicroCERCL, a root cause localization approach in a cloud-edge collaborative environment. Kernel network logs are extracted to detect kernel-level failures. After that, the application-level failure is further localized based on metric data. Based on failure backpropagation and cloud-edge topology aggregation, a graph neural network is designed to obtain the root cause ranking. Extensive experiments are conducted on datasets collected from the benchmark in a cloud-edge collaborative environment. MicroCERCL improves the accuracy significantly over the baselines while having better applicability.\par
In the future, we will extend MicroCERCL to improve the implementation to resolve the limitations mentioned in §5.2, further improve root cause localization accuracy, and ensure efficiency.

\bibliographystyle{ACM-Reference-Format}
\bibliography{sample-base}










\end{document}